\documentstyle[aas2pp4,psfig]{article}

\def\degree{{$^\circ$}}
\def\ts{\thinspace}           

\def\puncspace{\ifmmode\,\else{\ifcat.\C{\if.\C\else%
\if,\C\else\if?\C\else\if:\C\else\if;\C\else\if-\C\else%
\if)\C\else\if/\C\else\if]\C\else\if'\C%
\else\space\fi\fi\fi\fi\fi\fi\fi\fi\fi\fi}%
\else\if\empty\C\else\if\space\C\else\space\fi\fi\fi}\fi}%
\def\SP{\let\\=\empty\futurelet\C\puncspace}

\def\ee#1{\ifmmode {} \times 10^{#1} \else ${} \times 10^{#1}$\fi}
\def\sub#1{\ifmmode _{#1} \else $_{#1}$\fi}
\def\sup#1{\ifmmode ^{#1} \else $^{#1}$\fi}

\def\about{\ifmmode \sim \else {$\sim\,$}\fi}
\def\lta{\lesssim}
\def\gta{\gtrsim}

 \mathcode`*="002A   

\def\pd#1{{\partial\displaystyle #1}} 

\def\degree{{\ifmmode ^\circ \else $^\circ$\fi}}

\def\gcmsq{{\hbox{g\ts cm\sup2}}\SP}

\def\Hz{{\hbox{Hz}}\SP}

\def\kHz{{\hbox{kHz}}\SP}

\def\rps{{\hbox{rad\ts s\sup{-1}}}\SP}

\def\mdot{{\ifmmode \dot M \else {$\dot M$}\fi}}
\def\mdote{{\ifmmode \dot M_E \else {$\dot M_E$}\fi}}
\def\mdoti{{\ifmmode \dot M_i \else {$\dot M_i$}\fi}}
\def\msun{{\ifmmode M_\odot \else {$M_{\odot}$}\fi}}

\slugcomment{Submitted to ApJ}

\lefthead{Markovi\'{c} \& Lamb}
\righthead{Lense-Thirring Precession and QPOs}

\begin{document}
\title{Lense-Thirring Precession and QPOs in
X-Ray Binaries}

 \author{Dragoljub Markovi\'{c} and Frederick K.
Lamb\altaffilmark{1}}
 \affil{University of Illinois at
Urbana-Champaign, Department of Physics \\
 1110 W. Green Street, Urbana, IL 61801, USA }

 \altaffiltext{1}{Also, Department of Astronomy,
University of Illinois at Urbana-Champaign.}

 \begin{abstract}
 It has recently been suggested that
gravitomagnetic precession of the inner part of
the accretion disk, possibly driven by radiation
torques, may be responsible for some of the
quasi-periodic X-ray brightness oscillations
(QPOs) and other spectral features with
frequencies between 20 and 300~Hz observed in the
power spectra of some low-mass binary systems
containing accreting neutron stars and black hole
candidates. We have explored the free and driven
normal modes of geometrically thin disks in the
presence of gravitomagnetic and radiation warping
torques.

We have found a family of low-frequency
gravitomagnetic (LFGM) modes with precession
frequencies that range from the lowest frequency
allowed by the size of the disk up to a certain
critical frequency $\omega_{\rm crit}$, which is
$\sim$1~Hz for a compact object of solar mass.
The lowest-frequency (lowest-order) LFGM modes are
similar to the previously known radiation warping
modes, extend over much of the disk, and have
damping rates $\gta10$ times their precession
frequencies. The highest-frequency LFGM modes
are tightly wound spiral corrugations of the
disk that extend to $\sim$10 times its inner
radius and have damping rates $\gta10^3$ times
their precession frequencies. A radiation warping
torque can cause a few of the lowest-frequency
LFGM modes to grow with time, but even a strong
radiation warping torque has essentially no
effect on the LFGM modes with frequencies
$\gta10^{-4}$~Hz.

We have also discovered a second family of
high-frequency gravitomagnetic (HFGM) modes with
precession frequencies that range from
$\omega_{\rm crit}$ up to slightly less than the
gravitomagnetic precession frequency $\omega_{\rm
gm,i}$ of a particle at the inner edge of the
disk, which is 30~Hz if the disk extends inward
to the innermost stable circular orbit around a
2\msun\ compact object with dimensionless angular
momentum $cJ/GM^2 = 0.2$. The lowest-frequency
HFGM modes are very strongly damped and have
warp functions and precession frequencies very
similar to those of the highest-frequency LFGM
modes. In contrast, the highest-frequency
(lowest-order) HFGM modes are very localized
spiral corrugations of the inner disk and are
weakly damped, with $Q$ values $\sim$2--50.

We discuss the implications of our results for
the observability of Lense-Thirring precession
in X-ray binaries.
 \end{abstract}


\section{INTRODUCTION}

The possibility of observing gravitomagnetic
(Lense-Thirring) precession of accretion disks
around compact objects was considered early in
the study of accretion onto black holes and
neutron stars. The conclusion reached then
(Bardeen \& Petterson \markcite{Pett75}1975; see
also Petterson \markcite{Pett77a}1977a,
\markcite{Pett77b}1977b, \markcite{Pett78}1978;
Hatchett, Begelman, \& Sarazin
\markcite{HBS81}1981) was that the combined
action of gravitomagnetic and internal viscous
torques forces the disk flow into the rotation
equator of the compact object at $\sim$100
gravitational radii and holds it there, if the
outer disk is misaligned or the disk is driven by
time-independent torques. As a result, it has
been widely thought that gravitomagnetic
precession of the inner disk is unpromising as an
explanation for X-ray oscillations.

Recently Ipser (1996) has investigated `trapped'
(i.e., localized; see Kato \& Honma 1991)
pressure perturbations that warp the innermost
parts of accretion disks, found that the frame
dragging near a Kerr black hole causes precession
of such perturbations, and discussed their
possible relation to the low-frequency
quasi-periodic oscillations (QPOs) observed in
the X-ray brightness of black hole candidates
(see van der Klis 1995). The existence of these
modes depends on the presence of specific (but
not implausible) non-Keplerian angular velocity
profiles. Ipser's general relativistic analysis
did not take into account the viscosity of the
gas in the disk.

More recently, Stella \& Vietri
\markcite{SV98}(1998) have suggested that
gravitomagnetic precession of the inner disk may
be responsible for the broad bumps that are
observed between 20 and 60~Hz in the power
spectra of some of the accreting neutron stars
in low-mass binary systems called atoll sources
and for the $\sim$15--65~Hz horizontal branch
quasi-periodic oscillation (QPO) observed in the
X-ray brightness of the so-called Z sources (see
van der Klis \markcite{vdK95}1995). It has been
conjectured (e.g., Stella
\markcite{Stella97a}1997a,
\markcite{Stella97b}1997b) that the tilt of the
inner disk required for gravitomagnetic
precession could be caused either by the
radiation torque thought to produce radiation
warping of accretion disks (see Iping \&
Petterson \markcite{IP90}1990; Pringle
\markcite{Pringle96}1996; Maloney, Begelman, \&
Pringle \markcite{MBP96}1996; Maloney \&
Begelman \markcite{MB97}1997) or by forcing of
the inner disk by the neutron star's magnetic
field. Cui, Zhang, \& Chen
\markcite{CZC97}(1998) subsequently suggested
that the $\sim\,$6--300~Hz QPOs observed in the
X-ray brightness of the galactic black hole
candidates (Miyamoto et al.\
\markcite{Miyamoto91}1991; Morgan, Remillard, \&
Greiner \markcite{MRG97}1997; Remillard
\markcite{Rem97}1997) may be produced by 
gravitomagnetic precession. These suggestions
have renewed interest in the possibility of
observing gravitomagnetic precession near
compact objects.

In this paper we report calculations of the
properties of gravitomagnetic and radiation
warping modes of geometrically thin, Keplerian
$\alpha$ disks. Our work goes beyond previous
work in several ways. First, we have computed the
radiation warping mode functions and their
frequencies and growth rates up to very high
mode numbers. Our results for the tilt
functions and precession frequencies of the
low-order radiation warping modes agree well
with previous results (Maloney et al.\
\markcite{MBP96}1996). Second, we have explored
the time-dependent gravitomagnetic modes of the
disk, including both undriven (free) modes and
modes that are being driven at the inner
boundary. Third, we have investigated the
combined effects of gravitomagnetic and
radiation torques. Finally, we have considered
not only isothermal disks but also disks with a
variety of radial temperature structures, and
inner and outer boundary conditions appropriate
to a variety of astrophysical situations.

We have found a family of low-frequency
gravitomagnetic (LFGM) modes with precession
frequencies that range from the lowest frequency
allowed by the size of the disk up to a certain
critical frequency $\omega_{\rm crit}$, which is
$\sim$1~Hz for a compact object of solar mass.
The lowest-order (fundamental) LFGM mode is the
zero-frequency mode found earlier by Bardeen \&
Petterson (1975). The other low-order LFGM modes
are long-wavelength, precessing corrugations of
the outer disk similar to the previously known
low-order radiation warping modes and have
damping rates $\gta10$ times their precession
frequencies. The highest-frequency LFGM modes are
very tightly wound spiral corrugations of the
inner disk that extend only to $\sim$10 times the
inner radius of the disk and have damping rates
$\gta10^3$ times their precession frequencies. A
radiation warping torque makes all the low-order
modes in this family time-dependent and can cause
a few of the lowest-frequency LFGM modes to grow
with time, but even a strong radiation warping
torque has essentially no effect on the modes
with frequencies $\gta10^{-4}$~Hz.

We have also discovered a second family of
high-frequency gravitomagnetic (HFGM) modes with
precession frequencies that range from
$\omega_{\rm crit}$ up to slightly less than the
gravitomagnetic precession frequency $\omega_{\rm
gm,i}$ of a particle at the inner edge of the
disk. The highest possible mode frequency is the
gravitomagnetic precession frequency at the
innermost stable circular orbit, which is 30~Hz
for a 2\msun\ compact object with dimensionless
angular momentum $cJ/GM^2 = 0.2$. The dozen
highest-frequency (lowest-order) HFGM modes are
very localized spiral corrugations of the inner
disk, have precession frequencies that differ by
a factor of two, and are weakly damped, with $Q$
values $\sim\,$2-50. The lowest-frequency HFGM
modes have shapes very similar to the
highest-frequency LFGM modes and are very
strongly damped.

The `trapped' precessing modes discussed by Ipser
(1996; see also references therein) and the HFGM
modes discussed here are both localized near the
inner edge of the disk, but unlike the trapped
modes, the HFGM modes do not require rotation or
epicyclic frequency profiles with any special
shape. In fact, the shapes, precession
frequencies, and damping rates of the
modes in both the families we have studied depend
only weakly on whether the gravitational potential
used is the $1/r$ Newtonian potential or one of
several pseudopotentials designed to mimic the
steeper effective gravitational potentials of
general relativity (see, e.g. Nowak \& Wagoner
1993).

In \S~2 we state our assumptions, introduce the
differential equation that describes the normal
modes, explain the boundary conditions that we
use, and describe our method of solution. In
\S~3 we present and discuss the solutions of
the warp equation when only the gravitomagnetic
torque is included, when only the radiation
warping torque is included, and when both
torques are included. In \S~4 we summarize our
conclusions and discuss briefly the implications
of our results for the observability of
Lense-Thirring precession in X-ray binaries.

\section{ASSUMPTIONS AND METHOD}

In the work reported here we make several
approximations to simplify the calculations and
facilitate comparison with previous work. We
have explored the normal modes of disks in the
$1/r$ Newtonian gravitational potential and in
several pseudopotentials. We treat the disk flow
in the Newtonian approximation and include
gravitomagnetic effects only to lowest
post-Newtonian order. Therefore our treatment is
accurate only outside the radius of the
innermost stable circular orbit. We use an
$\alpha$ model of the accretion disk, setting
$T_{r\phi} = \alpha P$ with $\alpha$ constant,
and assume that the gas in the disk is fully
ionized hydrogen.

In considering the radiation-warping torque we
follow Pringle \markcite{Pringle96}(1996),
treating the central object as an isotropically
radiating point source, treating the disk as a
perfect absorber, and neglecting shadowing of the
disk at larger radii by the disk at smaller
radii. We also neglect the radial and drag forces
caused by interaction of radiation from the
central source with the gas orbiting in the disk
and the torques produced by radiation of the
kinetic energy dissipated within the disk (see
Miller \& Lamb \markcite{ML96}1996).

Following Petterson \markcite{Pett77a}(1977a),
we assume that the tilt of any given ring of
gas is small and treat the accretion disk as a
one-parameter family of rings that exchange
angular momentum via advection and viscous
torques. Writing the angular momentum per unit
radial distance and azimuthal angle as $\Sigma
R^3 \Omega {\bf l}$, where $\Sigma$ is the local
surface density of the disk, $R$ is the radius,
$\Omega$ is the angular velocity of the disk
flow, and ${\bf l}(R)$ is the unit vector
normal to the plane of the ring at $R$, the
equation that expresses conservation of angular
momentum is (compare Papaloizou \& Pringle
\markcite{PP83}1983, eq.~[2.4])
 \begin{eqnarray}
 \label{AngMomCons}
 \frac{\pd{}}{\pd t}
     \left(\Sigma R^3\Omega {\bf l}\right)
 \hspace{-0.2cm}  &+& \hspace{-0.2cm}
 \frac{\pd{}}{\pd R}
 \left(v_{\rm R} \Sigma R^3\Omega{\bf l}\right) 
 \nonumber \\ 
 &&\hspace{-2.0cm}
 = \frac{\pd{}}{\pd R}
     \left[
        \Sigma R^3
        \left(
            \nu_1\frac{\pd\Omega}{\pd R}{\bf l}
             + \frac{1}{2}\nu_2\Omega
               \frac{\pd{\bf l}}{\pd R}
       \right)
    \right]
 \nonumber \\
 &&\hspace{-1.5cm}
    + \frac{2G}{R^3 c^2}{\bf J}\times {\bf l}
           \left(\Sigma R^3\Omega\right)
    - s\,\frac{RL}{12\pi c}\hat{\bf z}
           \times\frac{\pd{\bf l}}{\pd R}\;,
 \end{eqnarray}
 where $\nu_1$ is the $R$-$\phi$ viscosity,
$\nu_2$ is the $R$-$z$ viscosity, $L$ is the
luminosity of the central source, and $\hat{\bf
z}$ and $J$ are the spin axis and angular
momentum of the star. The first term on the
right side of equation~(\ref{AngMomCons}) is
the viscous torque, the term proportional to
${\bf J} \times {\bf l}$ is the gravitomagnetic
torque produced by the spin of the compact
object, and the term proportional to $L$ is the
radiation torque. The factor $s$ in this term
allows us to track the effect of the radiation
torque: to include this torque we set $s=1$; to
neglect it, we set $s=0$.

In this report we are most interested in
gravitomagnetic precession of the disks around
the neutron stars in the atoll and Z sources.
These neutron stars are likely to have been spun
up to their equilibrium spin rates (Lamb et al.\
1985; Miller, Lamb, \& Psaltis 1998; Psaltis et
al.\ 1998). The $z$-component of the net flux of
angular momentum through the disk, $\dot J_z$,
is therefore zero. We therefore assume that the
$z$-component of the net flux (which is equal to
the component of the net flux parallel to the
axis of the disk, to first order in the tilt) is
zero.

In order to simplify the presentation, the
equations in the remainder of this section
assume that the angular velocity in the disk is
Keplerian in the $1/r$ Newtonian potential. (The
generalization to pseudopotentials is
straightforward.) 

For Keplerian disks with zero net angular
momentum flux along the $z$-axis, $v_{\rm R} =
(\nu_1/\Omega) (\pd\Omega/\pd R) =
-(3/2)(\nu_1/R)$ everywhere. To facilitate
comparison with previous work, we consider disks
with power-law radial temperature profiles $T(R)
= T_{\rm i}  (R/R_{\rm i})^{-\mu}$, where the
subscript i indicates the value of the quantity
at the inner radius of the calculation. In the
$\alpha$-disk model, $-\rho\nu_1 R(\pd\Omega/\pd
R) = 2\alpha\rho k_{\rm B}T/m_{\rm p}$ and hence
 \begin{equation}
 \label{vu1}
 \nu_1 =
 \frac{4\alpha k_{\rm B}T_{\rm i}}
      {3m_{\rm p}(G M)^{1/2}}
      R_{\rm i}^{\mu} R^{3/2-\mu}.
 \end{equation} 
 For simplicity, in the present work we take the
ratio $\kappa \equiv \nu_2/\nu_1$ of the two
kinematic viscosities to be independent of
radius.

For small tilt angles, ${\bf l} \approx
(\beta\cos\gamma, \beta\sin\gamma, 1)$ and hence
the disk warp is completely specified locally by
the tilt $\beta$ and twist $\gamma$. Following
Hatchett et al.\ \markcite{HBS81}(1981), we
represent the warp by the real and imaginary
parts of the complex variable $W \equiv
\beta\cos\gamma + i\beta\sin\gamma = \beta
e^{i\gamma}$. Then the component of
equation~(\ref{AngMomCons}) perpendicular to
$\hat{\bf z}$ can be written as (compare
Papaloizou \& Pringle \markcite{PP83}1983,
eq.~[2.6])
 \begin{eqnarray}
 \label{dWdt}
 \frac{\pd W}{\pd t} 
 \hspace{-0.2cm} &=& \hspace{-0.2cm}
     \frac{1}{2\Sigma R^3 \Omega}
     \frac{\pd{}}{\pd R}
     \left(\nu_2\Sigma R^3 \Omega
     \frac{\pd W}{\pd R}\right)
 \nonumber \\ 
 &&\hspace{0cm} + \hspace{0.1cm}
     i\frac{2GJ}{R^3c^2}W
   - is\Gamma\frac{\pd W}{\pd R}\; .
 \end{eqnarray}
 Here 
 \begin{equation}
 \label{Gamma}
 \Gamma \equiv
 L(12\pi\Sigma R^2\Omega c)^{-1} =
 G_0 R_{\rm i}^{\mu} R^{1-\mu} ,
 \end{equation}
 where
 \begin{eqnarray}   
 \label{Gzero}
  G_0 \hspace{-0.2cm} &\equiv &
  \hspace{-0.2cm}
  \frac{\epsilon\alpha c k_B T_{\rm i}}
  {3GMm_{\rm p}}
  \nonumber \\
        &=&\hspace{-0.2cm}
        3.1\ee{-4}\,\frac{\rm rad}{\rm s}
  \left(\frac{\epsilon^{}}{0.2_{}}\right)\!\!
  \left(\frac{\alpha^{}}{0.05_{}}\right)\!\!
  \left(\frac{T_{\rm i}}{10^7\,{\rm K}}\right)\!\!
  \left(\frac{2M_{\odot}}{M}\right)
  \qquad \nonumber \hspace{-1cm}\\
 \end{eqnarray}
 is the characteristic angular frequency
associated with the radiation torque at the
inner edge of the disk, expressed in terms of
the accretion efficiency $\epsilon$. In
writing the last expression on the right in
equation~(\ref{Gamma}), we have used the mass
continuity equation and the expression for the
radial velocity.

We now change to the new radial variable
 \begin{equation}
 \label{x.def}
 x \equiv
 ({\epsilon}/{\kappa})
 \left({R}/{R_{\rm g}}\right)^{1/2}\;,
 \end{equation}
 where $R_g \equiv GM/c^2$, and look for global
modes of the disk of the form $W(t,R) = e^{i\eta
t} W(x)$. Equation~(\ref{dWdt}) then becomes
 \begin{equation}
 \label{WarpEqn}
  xW'' + \left( 2 - isx \right)W'
 - i\!\left(A - \frac{B}{x^6}\right)\!\!
 \left(\frac{x}{x_{\rm i}}\right)^{\!\!2\mu}
 \!\! W
 = 0 , 
 \end{equation}
 where the prime denotes differentiation with
respect to $x$,
 \begin{equation}
 \label{defA}
  A \equiv 2\eta/G_0
 \end{equation}
 is the complex mode frequency scaled by $G_0$,
and
 \begin{eqnarray}
 \label{def.B}
 \hspace{-4cm}B &\equiv&
   \hspace{-0.2cm}
   j\frac{4\,c}{G_0 R_{\rm g}}
   \left(\frac{\epsilon^{}}
        {\kappa_{}}\right)^6
   \nonumber \\
 &=&\hspace{-0.2cm}
  \frac{1.7\ee{4}}{\kappa^6}
  \left(\frac{j}{0.2_{}}\right)
  \left(\frac{\epsilon^{}}
      {0.2_{}}\right)^{\!\!5}
  \left(\frac{0.05^{}}{\alpha_{}}\right)
  \!\left(\frac{10^7{\rm K}}{T_{\rm i}}\right)
 \end{eqnarray}
 is the gravitomagnetic frequency at $x=1$
scaled by $G_0$, for a compact object with
dimensionless angular momentum $j \equiv
cJ/GM^2$.

The strong dependence of $B$ on the accretion
efficiency $\epsilon$ and viscosity ratio
$\kappa$ is an artifact of our choice $x \propto
\epsilon/\kappa$ for fixed $R$. The
gravitomagnetic precession frequency at a given
radius $R$ is independent of $\epsilon$ and
$\kappa$, and the radius at which the
gravitomagnetic and radiation torques are
comparable therefore depends only weakly on
$\epsilon$ and $\kappa$. Note also that if
$s$ is set to zero, so that the radiation torque
is neglected, equation~(\ref{WarpEqn}) becomes
independent of $\epsilon$, as it must; in
this case $\epsilon$ is simply an arbitrary
factor in the conversion from $R$ to $x$ and has
no physical meaning.

Equation~(\ref{WarpEqn}) is a linear, complex,
second-order ordinary differential equation with
complex frequency $A$. Specifying a solution of
this equation, including the complex frequency,
therefore requires six conditions. The tilt
amplitude and twist angle at the inner edge of
the disk are arbitrary, and hence only four
conditions are physically meaningful. In
general, the physics imposes two boundary
conditions at the inner edge of the disk, at
$x_{\rm i}$, and two at the outer edge, at
$x_{\rm o}$. These boundary conditions restrict
solutions to a countable set of ``normal modes''
$\beta(x)$ and $\gamma(x)$ corresponding to a
discrete spectrum of precession frequencies
$\omega = {\rm Re}\,A$ and growth rates $\sigma
= -\,{\rm Im}\,A$.

Consider first the boundary conditions at the
inner edge of the disk. The solution depends
only on the ratio $W'(x_{\rm i})/W(x_{\rm i})$,
because $W(x_{\rm i})$ is arbitrary. We have
explored ratios from zero to extremely high
values. In particular, we have considered
$W'(x_{\rm i}) = 0$, which corresponds to no
external torque acting on the inner edge of the
disk. We have also considered a variety of
external torques that vary in time as $e^{i\eta
t}$ and act on the inner edge of the disk,
including a torque that forces $W(x_{\rm i}) =
0$. This boundary condition corresponds to
pinning of the inner edge of the disk in the
rotation equator of the compact object and is
only possible if there is a strong external
torque.

Consider now the boundary conditions at the
outer edge of the disk. If there is no torque on
the outer edge of the disk, then $W'(x_{\rm o})
= 0$. If instead the outer edge of the disk is
pinned to the equatorial plane, for example by
the accretion stream, the appropriate boundary
condition would be $W(x_{\rm o}) = 0$. For a
detailed discussion of outer boundary
conditions, see Maloney, Begelman, \& Nowak
(1998).

The warp equation~(\ref{WarpEqn}) is very stiff
when the gravitomagnetic torque is included. We
therefore used the method of Kaps and Rentrop
(see Press et al.\ \markcite{NumRec}1992,
\S~16.6) to integrate the warp equation outward
or inward, depending on the type of solution
being sought (see \S~3). In either case, we
varied the precession frequency and the growth
rate, iterating until the boundary condition at
the other edge of the disk from which we started
was satisfied. We have obtained approximate
analytical solutions of the warp equation in a
variety of asymptotic regimes, using the WKB
method. These solutions will be reported in
detail elsewhere (Markovi\'c \& Lamb, in
preparation). In many cases we used {\em
locally\/} valid analytic solutions to start or
guide the numerical integration. In some cases
(see below) we were only able to find global
numerical solutions after we had obtained locally
valid analytic solutions of the same type. We have
checked our global numerical solutions against
our analytical solutions wherever possible.

\section{RESULTS AND DISCUSSION}

We have solved the warp equation~(\ref{WarpEqn})
for the normal modes of the disk up to very high
mode numbers for a variety of gravitational
potentials, wide ranges of the relevant
parameters, the different inner and outer
boundary conditions discussed in the previous
section, and a range of outer disk radii.  In
this first report, we focus on our results for
the undriven (free) modes of isothermal disks in
the $1/r$ Newtonian gravitational potential,
mentioning only in passing our results for modes
driven at the inner edge of the disk and
pseudopotentials that mimic the steeper
effective potentials of general relativity.

We are particularly interested in
gravitomagnetic precession of the disks around
the neutron stars in the atoll and Z sources. As
noted in \S~2, these neutron stars are likely to
have been spun up to their equilibrium spin
rates, in which case $\dot J_z$, the
$z$-component of the net flux of angular
momentum through the disk, is zero. Therefore
all the results reported here are for $\dot J_z
= 0$. This assumption is not as restrictive as
it might seem, because the tilt functions,
precession frequencies, and damping rates of the
disk warping modes are almost unaffected by the
value of $\dot J_z$, unless it is almost equal
to the flux being advected through the disk.

When only the gravitomagnetic torque is
included, we have found two families of
solutions: a set of low-frequency
gravitomagnetic (LFGM) modes, all of which have
frequencies less than a certain critical
frequency $\omega_{\rm crit}$, and a set of
high-frequency gravitomagnetic (HFGM) modes, all
of which have frequencies greater than
$\omega_{\rm crit}$. Using approximate
analytical expressions valid for the high-order
modes of an isothermal disk, we find (see
Markovi\'{c} \& Lamb, in preparation)
 \begin{equation}
 \label{wcrit}
 \omega_{\rm crit} \approx
  (x_{\rm i}/x_{\rm o})^{1/2}B/(11 x_{\rm i}^6)\;,
 \end{equation}
 where $x_{\rm i}$ and $x_{\rm o}$ are the
dimensionless radii (see eq.~[\ref{x.def}]) of
the inner and outer edges of the disk and $B$ is
given by equation~(\ref{def.B}).

The critical frequency can be understood
qualitatively as follows. The precession
frequency of a gravitomagnetic mode is a
weighted average of the single-particle
precession frequency over the radial extent of
the mode, and hence $\omega_{\rm crit}$ is
roughly determined by the radius at which the
mode function peaks. The tilt functions of the
highest-order LFGM and HFGM modes all peak at
about the same place in the disk, so their
frequencies are almost identical. The value of
$x$ at which these modes peak scales with the
dimensions of the disk as
$x_i^{11/12}x_o^{1/12}$.

When only the radiation warping torque is
included, we have found a family of low-frequency
radiation-warping (R) modes. Our results for the
tilt functions and precession frequencies of the
low-order R modes agree well with previous
results (Maloney et al.\ 1996).
 When both the gravitomagnetic and
radiation-warping torques are included, we have
found a family of low-frequency gravitomagnetic
radia\-tion-warping (LFGMR) modes. The
lowest-order LFGMR modes are essentially R modes,
whereas the high-order LFGMR modes are
essentially LFGM modes. In between there are a few
hybrid modes that are affected by both the
radiation-warping torque and the gravitomagnetic
torque. The HFGM modes are essentially unaffected
by any radiation warping torque.

All the results reported here are for
$\kappa=1$, $G_0 = 3.1\ee{-4}$~\rps,
$B=1.7\ee{4}$, $x_{\rm i}=0.49$, and $x_{\rm
o}=50$. This value of $x_{\rm i}$ corresponds
to $R_{\rm i} = 6\,GM/c^2$ for $\epsilon=0.2$.
If the outer radius of the disk is two or three
times larger than assumed here, there would be a
few additional modes with very low frequencies.
The results shown for the R and HFGM modes assume
$W'(x_{\rm i}) = 0$ and $W'(x_{\rm o})=0$; the
results shown for the LFGM and LFGMR modes
assume only $W'(x_{\rm o})=0$.

Although in some cases the detailed shapes of the
warp functions and the precise values of the
frequencies and growth rates of the disk warping
modes depend on the choice of parameters and
boundary conditions, the general behavior of the
modes is very similar to the results we present
here. A detailed survey of parameters, boundary
conditions, and gravitational potentials will be
presented elsewhere (Markovi\'c \& Lamb, in
preparation).

\subsection{Low-Frequency Gravitomagnetic Modes}

The properties of the LFGM modes are determined
primarily by the gravitomagnetic and viscous
torques. These modes can therefore be found by
integrating the warp equation~(\ref{WarpEqn})
outward from the inner edge of the disk, using
the analytical solution that is valid at
small radii when the gravitomagnetic torque is
dominant to start the integration. (If instead
the warp equation is integrated inward, generally
only solutions with unphysically large torques at
the inner boundary can be found.)

 \begin{figure}[t!]    
 \vskip -0.14 truein
 \centerline{\hskip+3.5truein
\psfig{file=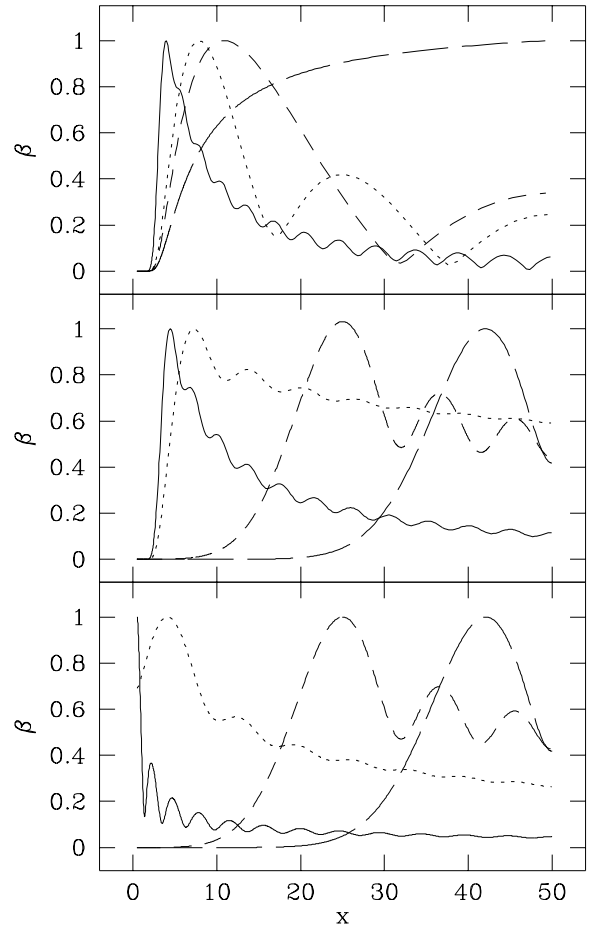,height=7.2 truein}}
 \vskip -2.0truein
 \caption[fig1]{
 Tilt functions $\beta(x)$ for selected warping
modes of the accretion disk, normalized to
maximum values of unity. 
 {\em Top\/}: Low-frequency gravitomagnetic
(LFGM) modes. Shown are the static $n=0$ mode
(long dashes) and the time-dependent $n=1$ (short
dashes), $n=2$ (dots), and $n=12$ (solid line)
modes.
 {\em Middle\/}: Low-frequency gravitomagnetic
radiation-warping (LFGMR) modes. Shown are the
time-dependent $n=1$ (long dashes), $n=3$ (short
dashes), $n=7$a (dots), and $n=12$ (solid line)
modes.
 {\em Bottom\/}: Radiation-warping (R) modes.
Shown are the time-dependent $n=1$ (long dashes),
$n=3$ (short dashes), $n=7$ (dots), and $n=12$
(solid line) modes.
 }
 \end{figure}

The LFGM warp functions, precession frequencies,
and damping rates are almost completely
independent of the inner boundary condition. In
particular, the LFGM modes do not depend on
whether the disk is being driven at the inner
boundary. The reason is that the gravitomagnetic
and viscous torques cause the tilt function to
converge to a characteristic shape that depends
only on the mode number within a radial distance
from the inner edge that is extremely small
compared to the radial extent of the mode,
regardless of what is happening at the inner
boundary. The properties of the low-order
precessing LFGM modes depend only weakly on
which boundary condition is imposed at $x_{\rm
o}$ and the value of $x_{\rm o}$, because even
these modes are large only near the inner edge
of the disk. The properties of the higher-order
LFGM modes are essentially independent of the
outer boundary condition and the radius of the
outer boundary. All LFGM modes are independent
of the accretion efficiency $\epsilon$.

The tilt functions $\beta(x)$ of four selected
LFGM modes are shown in the top panel of
Figure~1. We define the LFGM ``mode number'' $n$
loosely as the number of peaks in $\beta(x)$. The
fundamental ($n=0$) LFGM mode is static ($\omega
= \sigma = 0$). This mode reflects the tendency
of the gravitomagnetic and viscous torques to
align the axis of the inner disk with the spin
axis of the compact object (Bardeen \& Petterson
1975). Strictly speaking, for the $n=0$ mode $W'$
approaches zero at large $x$ only asymptotically;
however, $W'$ is already very close to zero at
$x=50$. All the time-dependent LFGM modes (those
with $n>0$) precess in the prograde direction
($\omega > 0$) and are damped ($\sigma < 0$).

Figure~2 shows the growth rates and precession
frequencies of the lowest $\sim$20 LFGM modes. As
$n$ increases, the relative spacing
$\Delta\omega/\omega$ between adjacent modes
decreases and the spectrum therefore becomes
quasi-continuous at large $n$. The spectrum of the
precession frequencies and damping rates of the
higher-order LFGM modes is plotted as a
continuous curve in Figure~3. The slight break in
the spectrum at $\omega \sim 10^2$ is where the
damping rate becomes comparable to $\omega_{\rm
crit}$.

 \begin{figure}[t!]   
 \vskip -0.43 truein
 \centerline{\hskip0.01truein
\psfig{file=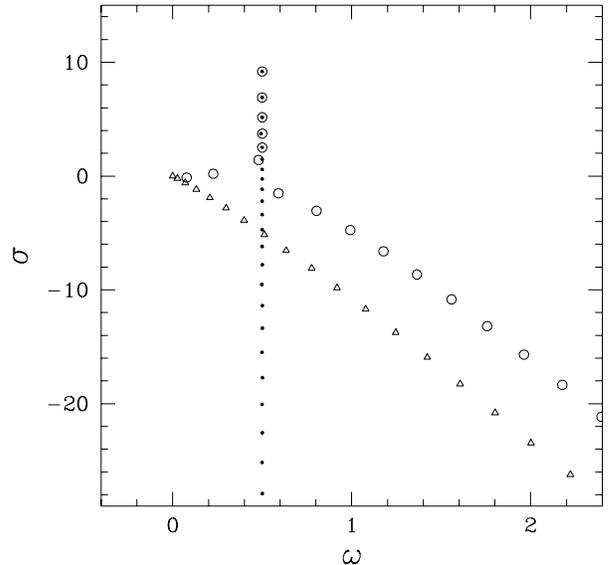,height=4.3truein}
 }
 \vskip -0.9 truein
 \caption[fig2]{
Spectrum of growth rates $\sigma$ and precession
frequencies $\omega$ of the lowest $\sim$20
low-frequency gravitomagnetic (LFGM) modes (open
triangles), radiation-warping (R) modes (dots),
and low-frequency gravitomagnetic
radiation-warping (LFGMR) modes (open circles) of
the accretion disk.
 }
 \end{figure}

The very high-order LFGM modes are spiral
corrugations of the inner disk that have very
short wavelengths and extend from the radius $
r_{\rm i}$ of the inner edge of the disk to
$\sim10r_{\rm i}$. Their tilt functions have a
single smooth peak, but the phase of the warp
changes by a very large amount in a very short
radial distance. When the radial distance
between corrugations becomes comparable to or
shorter than the vertical thickness of the disk,
treating the disk thickness as infinitesimal is
no longer valid. The dotted portion of the LFGM
spectrum shown in Figure~3 shows where this
distance becomes smaller than $10^{-2}\,r$. At
this point the precession frequency $\omega$ is
2\ee3, which corresponds to a circular frequency
$f\equiv {\rm Re}\,\eta/2\pi = \omega G_0/4\pi
\approx 0.05$~Hz.

 \begin{figure*}[t!]  
 \vskip -2.6truein
 \centerline{\hskip -0.4truein
\psfig{file=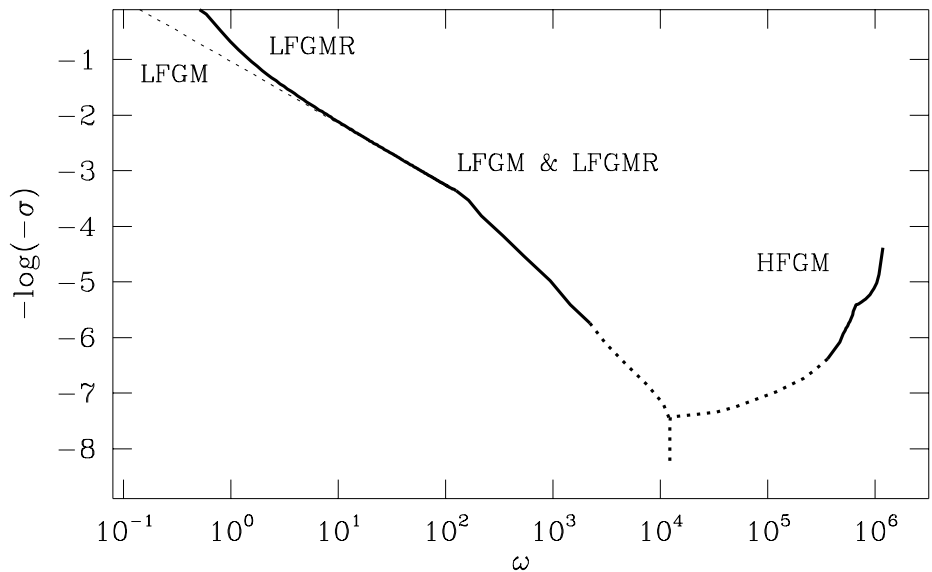,height=7truein}}
 \vskip -1.4truein
 \caption[fig3]{
 Spectrum of damping rates $-\sigma$ and
precession frequencies $\omega$ of low-frequency
gravitomagnetic (LFGM) modes (dotted line),
low-frequency gravitomagnetic radiation-warping
(LFGMR) modes (left solid line), and
high-frequency gravitomagnetic (HFGM) modes
(right solid line) plotted as continuous curves.
 The damping rates of the LFGM and LFGMR modes
with $\omega$ greater than $\sim$10 are
indistinguishable on this scale. The LFGM and
HFGM spectra become almost vertical at the
critical frequency $\omega_{\rm crit}$ discussed
in the text.
 The HFGM spectrum is essentially unaffected by
any radiation warping torque.
 The dotted portions of the curves show where the
radial distance between corrugations becomes
smaller than $10^{-2}\,r$.
 }
 \end{figure*}

The precession frequencies of the LFGM modes
increase with $n$, but only up to $\omega_{\rm
crit} \sim 1.2\ee4$ for $\epsilon =0.2$ and
$x_{\rm i}=0.49$. As $n$ increases further,
$\omega$ remains $\sim 1.2\ee4$, which
corresponds to a circular frequency $\sim\,1$~Hz
for a compact object of solar mass. The reason is
that as $n$ increases, the LFGM modes at first
become more localized near the inner edge of the
disk but eventually stop becoming localized.
Hence, the precession frequency stops rising.
This is the case not only for the free LFGM modes
but also for LFGM modes driven at the inner
boundary and for the LFGM modes in
pseudopotentials that mimic the effective
potential of general relativity. Thus, {\em all
LFGM modes have precession frequencies
$\lta1\,$Hz for compact objects of solar mass.}

The damping rates of even the lowest-order
precessing LFGM modes are $\gta10$ times higher
than their precession frequencies. For example,
the precession frequencies and growth rates of
the LFGM modes shown in Figure~1 are
$\omega=0.026$, $\sigma =-0.209$ for $n=1$;
$\omega=0.071$, $\sigma = -0.599$ for $n=2$; and
$\omega=1.25$, $\sigma = -13.7$ for $n=12$. As
$n$ increases, the damping rate increases faster
than the precession frequency (see Fig.~2). Thus,
modes with precession frequencies $\omega \approx
100$ have damping ratios $|\sigma|/\omega$
$\gta20$. At the critical frequency ($\omega_{\rm
crit} \sim 1.2\ee4$), the damping ratio exceeds
$1.5\ee3$. Hence, {\em all precessing LFGM modes
are highly overdamped}.

\subsection{Radiation Warping Modes}

It has been suggested (e.g., Stella
\markcite{Stella97a}1997a,
\markcite{Stella97b}1997b) that the radiation
warping torque discussed by Pringle
\markcite{Pringle96}(1996) and Maloney et al.\
\markcite{MBP96}(1996) may drive gravitomagnetic
precession modes of the inner disk. We have
investigated the effects of this torque on the
precession frequencies and damping rates of
gravitomagnetic modes.

We first discuss the normal modes of the disk
when the radiation-warping torque is included
but the gravitomagnetic torque is not. These
modes can be readily found by numerically
integrating the warp equation~(\ref{WarpEqn})
outward from the inner boundary or inward from
the outer boundary, because the warp equation is
not stiff when the gravitomagnetic torque is
neglected.

The detailed shapes of the low-order R-mode tilt
functions are affected by which boundary
condition is imposed at the outer edge of the
disk at $x_{\rm o}$ and by the value of $x_{\rm
o}$. In contrast, the tilt functions of the
high-order R modes are strongly concentrated near
the inner edge of the disk and hence are almost
unaffected by which boundary condition is imposed
at $x_{\rm o}$ or the value of $x_{\rm o}$. The
high-order R-mode tilt functions typically become
steep very near the inner boundary at $x_{\rm
i}$, even though $\beta'(x_{\rm i})=0$, because
$\beta''(x_{\rm i})$ is large.

Although the tilt functions of the high-order R
modes are almost unaffected by which boundary
condition is imposed at $x_{\rm o}$ or the value
of $x_{\rm o}$, the precession frequencies of
{\em all\/} the R modes depend on which boundary
condition is imposed at the outer boundary. 

The tilt functions $\beta(x)$ of four R modes are
shown in the bottom panel of Figure~1. We find
that the tilt functions of the lowest-order R
modes are appreciable only at large radii, in
agreement with previous work (Maloney et al.\
\markcite{MBP96}1996). The precession
frequencies of the half-dozen lowest-order R
modes are therefore independent of the inner
boundary condition.

The dots in Figure~2 show the growth rates and
precession frequencies of the lowest $\sim\,$20 R
modes. We find that all R modes that satisfy
$W'(x_{\rm i})=W'(x_{\rm o})=0$ precess in the
prograde direction with frequency $\omega =
0.5$. The growth rates of the four low-order R
modes shown in Figure~1 are $\sigma = 9.19$ for
$n=1$, which is the lowest-order R mode; $\sigma
= 5.16$ for $n=3$; $\sigma = 0.58$ for $n=7$;
and $\sigma = -4.72$ for $n=12$. The growth rate
falls steeply with increasing mode number, and
modes with $n \gta 10$ are very strongly damped.
For fixed mode number $n$, the growth rate
increases with increasing $x_{\rm o}$, i.e.,
$d\sigma/dx_{\rm o} > 0$.

We also find that all the R modes that satisfy
$W(x_{\rm i})=W(x_{\rm o})=0$ precess in the
prograde direction with frequency $\omega = 1$.
This is the same precession frequency found by
Maloney et al.\ (1996) for all the R modes that
satisfy $x^2 W'(x) \rightarrow 0$ as
$x\rightarrow 0$ instead of $W'(x_{\rm i}) = 0$.
This degeneracy of R-mode precession frequencies
is peculiar to isothermal disks: for more
general temperature profiles, different R modes
precess at different rates (see Maloney et al.\
1998). On the other hand, for R modes that
satisfy ``mixed'' boundary conditions, either
$W(x_{\rm i})=W'(x_{\rm o})=0$ or $W'(x_{\rm
i})=W(x_{\rm o})=0$, the higher-order modes
precess at slightly different frequencies from
the lowest-order modes. For $W(x_{\rm
i})=W'(x_{\rm o})=0$, most modes precess in the
prograde direction, but the torque acting on the
inner edge of the disk causes the tilt function
of one of the low-order modes to increase
monotonically with increasing $x$, i.e.,
$\beta'(x)$ is positive everywhere. This mode
precesses very slowly, either in the prograde or
the retrograde sense, depending on the value of
$x_{\rm o}$, and is either weakly growing or
weakly damped (see Markovi\'{c} \& Lamb, in
preparation).

\subsection{Low-Frequency Gravitomagnetic
Radiation Warping Modes}

Consider now the low-frequency normal modes of the
disk when both the gravitomagnetic torque and the
radiation warping torque are included. These are
the LFGMR modes. These modes can be found by
integrating the warp equation~(\ref{WarpEqn})
outward from the inner edge of the disk, using
the same approach that was used to find the LFGM
modes.

The character of the LFGMR modes changes with
increasing mode number. Like the tilt functions
of the low-order R modes, the tilt functions of
the low-order LFGMR modes are appreciable only
at large radii and are affected by which
boundary condition is imposed at $x_{\rm o}$ and
by the value of $x_{\rm o}$. Like the precession
frequencies of all the R modes, the precession
frequencies of the low-order LFGMR modes depend
on which boundary condition is imposed at the
outer boundary, but not on the value of $x_{\rm
o}$. Unlike the R modes but like the LFGM modes,
the frequencies and damping rates of the LFGMR
modes are almost completely independent of the
inner boundary condition, for the same reason
that this is true of the LFGM modes. The
properties of the high-order LFGMR modes are
only weakly dependent on the boundary condition
at $x_{\rm o}$ and the value of $x_{\rm o}$.
Only the properties of the lowest-order LFGMR
modes depend on the accretion efficiency
$\epsilon$.

For the parameter values considered in this
report, the first six LFGMR modes are appreciable
only at large radii and hence are virtually
unaffected by the gravitomagnetic torque. Their
tilt functions, precession frequencies, and
growth rates are essentially identical to those
of the first six R modes (see Figs.~1 and~2),
except that the very innermost part of the disk
is forced into the rotation equator of the
central object by the combined action of the
gravitomagnetic and internal viscous torques, as
first discussed by Bardeen \& Petterson (1975).
This is the reason that the properties of the
lowest-order LFGMR modes do not depend on the
boundary condition at the inner edge of the disk. 

Like the low-order radiation warping modes, the
lowest-order LFGMR modes precess in the prograde
direction with precession frequencies of either
0.5 or 1.0, depending on whether the inner and
outer edges of the disk are free or pinned in
the disk plane. For a compact object of solar
mass, these precession frequencies are $\sim
10^{-5}\,$Hz. Also like the low-order radiation
warping modes, the lowest-order LFGMR modes are
all growing modes.

There are typically two or three LFGMR modes that
are hybrid gravitomagnetic radiation-warping
modes. The shapes of these modes are very similar
to the radiation-warping modes in the outer
disk,  except that the combined action of the
gravitomagnetic and internal viscous torques
again forces the very innermost part of the disk
into the rotation equator of the central object.
The hybrid gravitomagnetic radiation-warping
modes generally (but not always; see below)
precess in the prograde direction with
precession frequencies between $\sim10^{-6}$~Hz,
for a compact object of solar mass ($\sim$0.08
in our dimensionless units), and the $\sim
10^{-5}$~Hz frequency of the radiation-warping
modes. These modes are either weakly damped or
weakly growing. Thus, when the radiation-warping
torque is taken into account, there is generally
no time-independent mode: the Bardeen-Petterson
mode becomes weakly time-dependent.

For some values of $x_{\rm o}$ there is a single
mode that precesses slowly in the {\em
retrograde\/} direction and is either weakly
growing or weakly damped. For this mode, the
gravitomagnetic torque plays a role analogous to
that of the inner boundary torque for R modes
with $W(x_{\rm i})=W'(x_{\rm o})=0$, pinning the
disk to the rotation equator of the central
object. For a special choice of the parameters
$B$ and $x_{\rm o}$, there is a single static
mode ($\omega = \sigma = 0$). However, because
of the fine tuning required, this situation is
unlikely to be of any practical importance.

The tilt functions of four low-order LFGMR modes
are shown in the middle panel of Figure~1. The
frequencies and growth rates of these modes are
$\omega=0.5$, $\sigma = 9.19$ for $n=1$, which is
the lowest-order LFGMR mode; $\omega=0.5$, $\sigma
= 5.16$ for $n=3$; $\omega=0.23$, $\sigma = 0.20$
for $n=7$a (see below); and $\omega=1.18$, $\sigma = -6.62$
for $n=12$. The growth rates and precession
frequencies of the lowest $\sim$20 LFGMR modes
are shown in Figure~2.

For the parameter values used in the present
study, there are two LFGMR modes that have tilt
functions with 7 peaks. The tilt function,
precession frequency, and growth rate of the
$n=7$a LFGMR mode shown in Figure~1 are
intermediate between those of the $n=7$ R mode
and the static LFGM mode. This is a growing mode
that precesses in the prograde direction with a
very low frequency and grows only very slowly
(see Fig.~2). There is also another LFGMR mode
with seven peaks, which we denote 7b; it
precesses in the prograde direction with an even
lower frequency ($\omega = 0.08$; again see
Fig.~2) and is very weakly damped ($\sigma =
-0.12$). This LFGMR mode is the static
gravitomagnetic mode, weakly perturbed by the
radiation torque.

All LFGMR modes with $n>7$ are damped, but less so
than LFGM modes with about the same frequency.
These LFGMR modes are only weakly affected by the
boundary condition at $x_{\rm o}$ or the value of
$x_{\rm o}$. The properties of the high-order
LFGMR modes are almost identical to those of the
LFGM modes with the same frequency. In particular,
{\em all but the lowest-order LFGMR modes are
highly overdamped\/} (the damping ratios of modes
with $\omega\gta5$ are $\gta10$). Also, with
increasing mode number, $\omega$ asymptotically
approaches $\sim1.2\ee4$ for a disk inner edge at
$x_{\rm i}=0.49$ (see Fig.~3). Again, this result
holds even if the disk is driven at its inner
edge and even for potentials that are steeper
than the $1/r$ Newtonian potential. Thus, {\em all
LFGMR modes have precession frequencies
$\lta1$~Hz for compact objects of solar mass}.

\subsection{High--Frequency Gravitomagnetic
Modes}

The properties of the HFGM modes, which are all
confined very near the inner edge of the disk, are
determined primarily by the gravitomagnetic
torque. If the warp equation is integrated
outward, generally only the LFGM modes can be
found, as noted in \S~3.1. Although the HFGM
modes can in principle be found by integrating
the warp equation~(\ref{WarpEqn}) inward from the
outer edge of the disk, this is generally
impractical. The reason is threefold. First, the
warp functions of the HFGM modes increase by a
factor of up to $\sim10^{7,000}$ in going from
the outer edge of the disk to their peak values,
which occur very near the inner edge. Second,
solving for the mode functions requires
searching the two-dimensional $\omega$-$\sigma$
space for the precise complex frequencies that
give valid solutions of the warp equation with
the boundary conditions chosen. The signature of
a valid solution is a mode function that is
well-behaved at the inner edge of the disk, but
because even valid solutions grow very large as
the integration proceeds, valid and invalid
solutions are not easily distinguishable without
integrating the warp equation all the way to the
inner edge of the disk, which is computationally
inefficient. Third, the warp equation is very
stiff at the high frequencies of these modes, so
searching numerically for the complex
frequencies of the valid solutions is difficult.

The lowest-order (highest-frequency) HFGM modes
were initially found by using analytical
solutions valid for large $\omega$ to start the
inward numerical integration at $x \approx 0.6$,
i.e., at an $x$ value $\sim$20\% larger than the
$x$ value at the inner edge of the disk. Starting
at this $x$ value, the HFGM mode functions
typically increase by only a factor of
$10^{50}$--$10^{100}$ between the starting point
and their peak, so the search for solutions in
the $\omega$-$\sigma$ plane is much more
efficient. Of course, once the complex frequency
of a solution is accurately known, the warp
equation can be integrated all the way from the
outer edge of the disk to its inner edge.
Moreover, once one valid solution has been
found, additional solutions can be found
relatively easily.

Unlike the LFGM and LFGMR modes, the warp
functions, precession frequencies, and damping
rates of the HFGM modes are affected by the
boundary condition at the inner edge of the disk.
In particular, the properties of the HFGM modes
depend on whether the disk is being driven at
its inner boundary. A strong driving torque at
the inner edge of the disk can excite a variety
of HFGM modes, depending on its time dependence
and components parallel and perpendicular to the
angular momentum of the gas in the disk. A
driving torque that oscillates with any given
frequency can excite a driven HFGM mode with
that same frequency, if the torque is strong
enough.

The frequencies and damping rates of the HFGM
modes are only slightly affected if the Newtonian
gravitational potential is replaced by a
pseudopotential that mimics the steeper effective
potentials of general relativity. The HFGM modes
are almost independent of the outer boundary
condition and the radius of the outer boundary.
They are essentially unaffected by the radiation
warping torque and hence do not depend on the
accretion efficiency $\epsilon$.

Here we focus on the undriven (free) HFGM modes,
which are the relevant HFGM modes when there is
no torque acting on either the inner or the outer
edge of the disk. These HFGM modes satisfy
$W'(x_{\rm i}) = 0$ and $W'(x_{\rm o}) =0$. All
the free HFGM modes are time-dependent, precess
in the prograde direction ($\omega > 0$), and are
damped ($\sigma < 0$).

We define the HFGM mode number $m$ by counting
downward in frequency, starting with $m=1$.  The
tilt functions $\beta(x)$ of four low-order free
HFGM modes are shown in Figure~4. All have a
single very narrow peak near the inner edge of
the disk. These modes are very localized, tightly
wound spiral corrugations of the disk very near
its inner edge. The radial distance between the
corrugations of the lowest dozen or so HFGM
modes decreases with increasing mode number,
although not smoothly, whereas the overall radial
width of the tilt function increases with
increasing mode number, approximately linearly
for $m>2$ up to $m \sim 14$.

Even though the HFGM warp functions are tightly
wound, the two lowest-order HFGM modes have
relatively small winding numbers $N \equiv
|\Delta\gamma|/2\pi$, where $\Delta\gamma$ is
the change in phase over the radial interval
where the mode has appreciable amplitude: $N$ is
$\sim0.5$ for $m=1$ and $\sim1$ for $m=2$. Hence
these two modes disturb significantly the
azimuthal symmetry of the disk. In contrast, the
high-order HFGM modes have high winding numbers
and are therefore almost axisymmetric. This is
illustrated in Figure~5, which shows the real and
imaginary parts of the warp functions for the
$m=1$, $m=2$, and $m=8$ HFGM modes.

 \begin{figure}[t!]  
 \vskip -0.3 truein
 \centerline{\hskip+3.7truein
\psfig{file=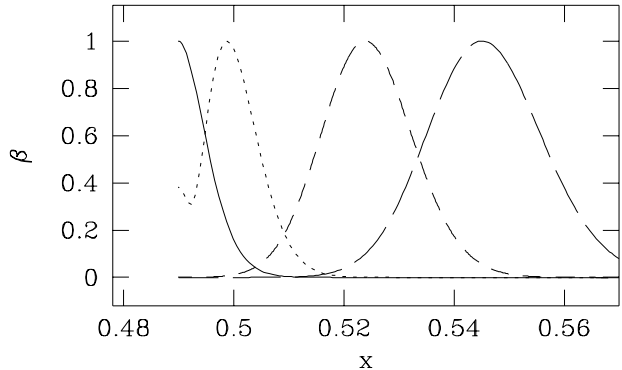,height=7.5truein}
 }
 \vskip -5.25truein
 \caption[fig4]{
 Tilt functions $\beta(x)$ of four low-order
undriven high-frequency gravitomagnetic (HFGM)
modes, normalized to maximum values of unity.
Shown are the time-dependent $m=1$ (solid line),
$m=2$ (dots), $m=8$ (short dashes) and $m=14$
(long dashes) modes.
 }
 \end{figure}

The lowest-order free HFGM modes precess with
frequencies comparable to the gravitomagnetic
precession frequency $\omega_{\rm gm,i} =
B/x_{\rm i}^6$ of a particle at the inner edge of
the disk. The highest possible precession
frequency is therefore the gravitomagnetic
precession frequency at the innermost stable
circular orbit. The lowest-order free HFGM modes
are weakly damped.

 \begin{figure*}[t!]   
 \vskip -0.2 truein
 \centerline{\hskip -0.1truein
 \psfig{file=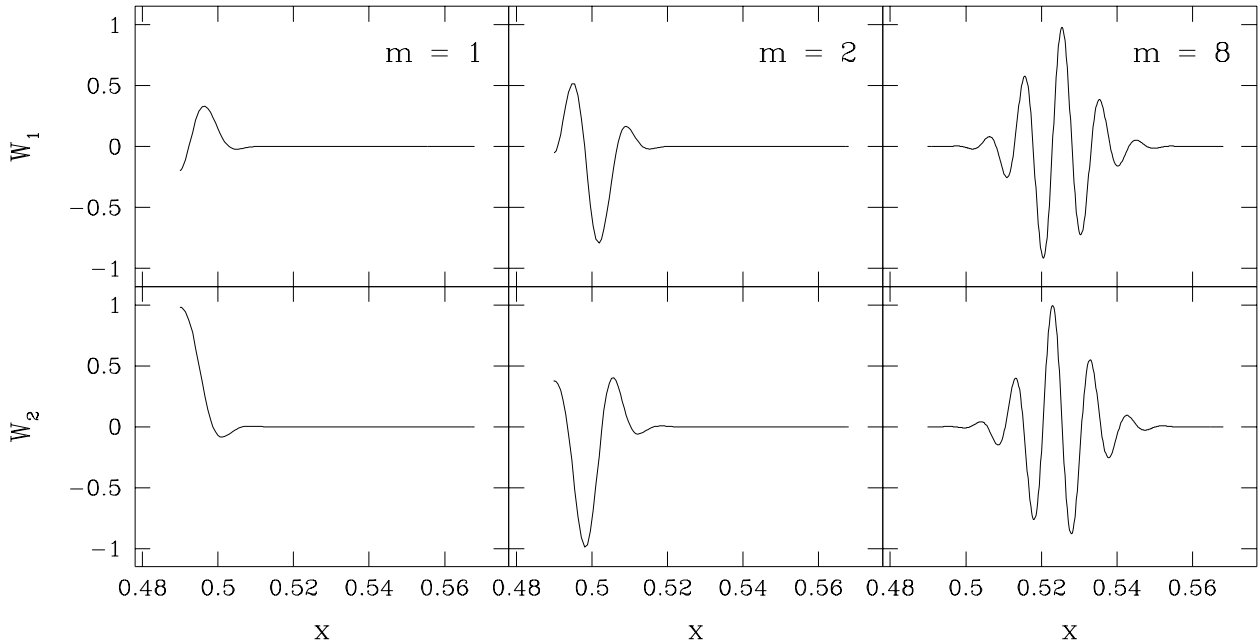,height=7.0 truein}}
 \vskip -3.4 truein
 \caption[fig3]{
 Real and imaginary parts $W_1(x)$ and $W_2(x)$
of the tilt function $W$ for the $m=1$, $m=2$,
and $m=8$ HFGM modes. The tilt functions
$\beta(x)$ of these three modes are shown in
Figure~4. 
 }
 \end{figure*}

Specifically, the precession frequencies and
growth rates of the first few HFGM modes are
$\omega = 1.19\ee6 = 0.97\,\omega_{\rm gm,i}$,
$\sigma = -2.40\ee4$ for $m=1$; $\omega =
1.10\ee6$, $\sigma = -7.32\ee4$ for $m=2$;
$\omega = 1.04\ee6$, $\sigma = -1.05\ee5$ for
$m=3$; $\omega = 8.22\ee5$, $\sigma = -2.02\ee5$
for $m=8$; and $\omega = 6.53\ee5 =
0.53\,\omega_{\rm gm,i}$, $\sigma = -2.87\ee5$
for $m=14$. The $Q$ ($\equiv -\omega/\sigma$)
values of these modes range from 49 for the $m=1$
mode to 2.3 for the $m=14$ mode. The damping
rates of these modes increase with increasing
viscosity. For example, if the viscosity
parameter $\alpha$ (see \S~2) is 0.5 rather than
0.05, the $Q$ of the fundamental HFGM mode
becomes 22 rather than 49.

Figure~6 shows the individual growth rates and
precession frequencies of the eleven lowest-order
(highest-frequency) free HFGM modes; the growth
rates and precession frequencies of the next
higher-order HFGM modes are indicated by the
dotted curve. The frequencies of the HFGM modes
at first decrease steadily with increasing mode
number, but the separation between adjacent modes
in the $\omega,\sigma$ plane shrinks rapidly with
further increases in the mode number and the
spectrum becomes quasi-continuous.

The spectrum of damping rates and precession
frequencies of the free HFGM modes up to very
high order are plotted as a continuous curve in
Figure~3, together with the spectrum of the free
LFGM and LFGMR modes for comparison. The dotted
portion of the HFGM spectrum shown in Figure~3
shows where the radial distance between
corrugations becomes smaller than $10^{-2}\,r$.
The corresponding precession frequency $\omega$ is
4\ee5, which is 10~Hz.

As Figure~3 shows, the damping rate increases
steadily with increasing mode number. At $m \sim
15$ ($\omega \approx 6.3\ee5$), $Q$ has fallen to
about unity ($\sigma \approx -1.1\ee6$). The
$\omega$-$\sigma$ spectrum breaks at this
frequency, because the damping rates of the
higher-order modes increase even more rapidly
with decreasing $\omega$ as the mode functions
broaden and their winding numbers increase.

As the mode number increases further and the
precession frequency approaches $\omega_{\rm crit}$
from above, the spectrum turns sharply downward,
because the damping rate continues to grow
whereas the frequency approaches the constant
critical frequency $\omega_{\rm crit}$ from
above. In the limit \hbox{$\omega \rightarrow
\omega_{\rm crit}$}, \hbox{$\sigma \rightarrow
-\infty$}, the tilt functions of the HFGM modes
assume the same shape $\beta_{\rm crit}(x)$
(which peaks near $x=1$) assumed by the tilt
functions of the LFGM modes as \hbox{$\omega
\rightarrow \omega_{\rm crit}$} from below.
Hence, {\em all the HFGM modes have precession
frequencies between $\sim\,1\,$Hz and
$\sim\,30\,$Hz for compact objects of solar
mass.} Although the high-order free HFGM modes
are very strongly damped, {\em the lowest-order
HFGM modes are only weakly damped}.

\section{SUMMARY AND CONCLUSIONS}

We have found two families of disk warping modes
in the presence of gravitomagnetic and
radiation-warping torques. In general, all of
these modes are time-dependent. In particular,
the static mode found by Bardeen \& Petterson
(1975) is perturbed by the radiation-warping
torque and becomes weakly time-depen\-dent.

The low-frequency gravitomagnetic
ra\-dia\-tion-warp\-ing (LFGMR) modes have
precession frequencies that range from the
lowest frequency allowed by the size of the
disk, which in dimensionless units (see \S~2) is
$\sim\,$0.1 for the parameter values used in the
present study, up to the critical frequency
$\omega_{\rm crit}$ (see eq.~[\ref{wcrit}]),
which is $\sim1.2\ee{4}$. The high-frequency
gravitomagnetic (HFGM) modes have precession
frequencies that range from $\omega_{\rm crit}$
up to slightly less than the gravitomagnetic
precession frequency $\omega_{\rm gm,i}$ of a
particle at the inner edge of the disk, which is
$\sim1.2\ee{6}$ for the parameter values used
in the numerical computations reported here. The
HFGM modes are essentially unaffected by any
radiation-warping torque.

The damping rates and frequencies of the modes in
both families are changed by $\lta10$\% if the
$1/r$ Newtonian potential is replaced by any of
several pseudopotentials that mimic the steeper
effective gravitational potentials of general
relativity, because the properties of the
modes at the inner edge of the disk are
determined primarily by the gravitomagnetic
torque.

The vast majority of the modes that lie between
the lowest-frequency LFGMR modes and the
highest-frequency HFGM modes are very
short-wavelength corrugations of the disk and are
very highly overdamped. In particular, the modes
with frequencies in the interval \hbox{$2\ee3 <
\omega < 4\ee5$} have corrugation wavelengths
$\lta$1\% of the radius at which the mode peaks.
When the wavelength becomes less than the
vertical thickness of the disk, our treatment of
the disk as infinitesimal is no longer valid
(see, e.g., Papaloizou \& Pringle 1983). In this
regime, modes of the type we have found may be
even more strongly damped or even nonexistent.

\subsection{The LFGMR Modes}

The half-dozen lowest-order
(lowest-frequency)\break LFGMR modes have
essentially the same shapes as the previously
known low-order radiation warping modes (Maloney
et al.\ 1996), except that the very innermost
part of the disk is forced into the rotation
equator of the central object by the combined
action of the gravitomagnetic and internal
viscous torques, as first discussed by Bardeen
\& Petterson (1975). This is the reason that,
unlike the radiation-warping modes, the
low-order LFGMR modes do not depend on the
boundary condition at the inner edge of the
disk.

 \begin{figure}[t!]   
 \vskip -0.5 truein
 \centerline{\hskip+0.0truein
\psfig{file=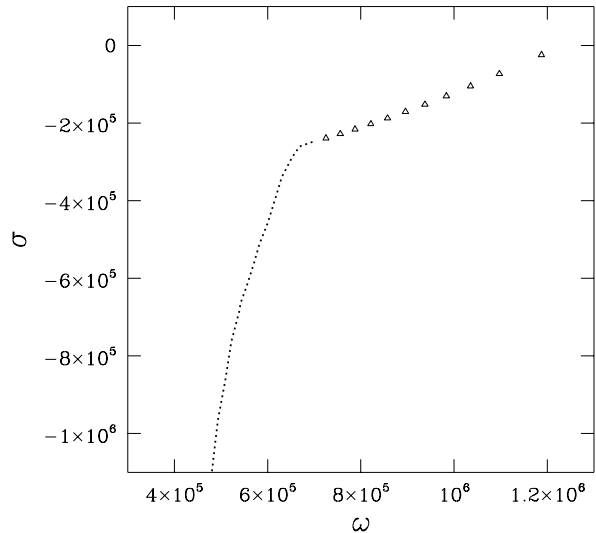,height=4.0truein}
 }
 \vskip -0.8truein
 \caption[fig4]{
 Spectrum of growth rates $\sigma$ and precession
frequencies $\omega$ of undriven high-frequency
gravitomagnetic (HFGM) modes of the accretion
disk. The frequencies and damping rates of the
eleven lowest-order (highest-frequency) HFGM
modes are plotted individually as open triangles.
The dense spectrum formed by the precession
frequencies and damping rates of the
higher-order HFGM modes is indicated by the
dotted line.
 }
 \end{figure}

The lowest-order LFGMR modes precess in the
prograde direction with precession frequencies
of either 0.5 or 1.0 in dimensionless units,
depending on whether the outer edge of the disk
is free or pinned in the disk plane. This is the
same precession behavior displayed by the
low-order radiation warping modes. For a compact
object of solar mass, these modes precess with
frequencies $\sim 10^{-5}$~Hz. The lowest-order
LFGMR modes, like the low-order radiation
warping modes, are growing modes.

Typically two or three low-order LFGMR modes have
properties intermediate between the
gravitomagnetic modes and the radiation-warping
modes. These hybrid modes are similar in shape
to the radiation-warping modes of the same
order, except that their tilt functions go to
zero at the inner edge of the disk because of
the combined action of the gravitomagnetic and
internal viscous torques. The hybrid modes
generally (but not always) precess in the
prograde direction and are either weakly damped
or weakly growing. For a compact object of solar
mass, the precession frequencies of these modes
lie between $\sim10^{-6}$~Hz and $\sim
10^{-5}\,$Hz, which is the frequency of the
lowest-order LFGMR modes.

With increasing mode number, the frequencies of
the LFGMR modes at first increase steadily but
eventually approach $\omega_{\rm crit}$
asymptotically from below. This happens even if
the disk is driven at its inner edge. The LFGMR
modes that have precession frequencies
$\gta10^{-4}$~Hz are essentially unaffected by
even a strong radiation-warping torque and have
damping rates $\gta10$ times their frequencies.
The highest-frequency LFGMR modes are very
short-wavelength spiral corrugations that extend
from the radius $r_{\rm i}$ of the inner edge of
the disk only to $\sim10\,r_{\rm i}$. These modes
have precession frequencies $\sim \omega_{\rm
crit}$, which is $\sim\,1$~Hz for a compact
object of solar mass, and are very strongly
damped.

The hybrid gravitomagnetic radiation-warping
modes that are excited by the radiation-warping
torque may be of interest observationally, if
they grow to significant amplitudes. These modes
have very low frequencies, as much as an order of
magnitude lower than the lowest-frequency modes
known previously.

The high-frequency LFGMR modes appear unpromising
as an explanation of the 6--300~Hz QPOs observed
in some accreting neutron stars and black hole
candidates, both because they are very strongly
damped and because they have frequencies
$\lta1$~Hz for compact objects of solar mass.
Excitation of high-frequency LFGMR modes by
impulsive disturbances or a broad spectrum of
fluctuations is unlikely to produce peaks in the
brightness power spectrum with relative widths
$\delta\nu/\nu$ as small as the $\sim\,$0.5--1
widths of the broad bumps observed in the
frequency range $\sim\,$20--40~Hz in some atoll
sources, let alone as small as the
$\sim\,$0.1--0.3 widths of the QPO peaks observed
in the Z sources and a several black-hole
candidates. A narrow peak could be created by
exciting these modes with a nearly monochromatic
driving force, but then the QPO would not be
generated by Lense-Thirring precession but by
the driving force and would have a frequency and
coherence unrelated to the properties of the
gravitomagnetic modes of the inner disk.
Moreover, the amplitudes of these modes would be
limited by their very high damping rates.

\subsection{The HFGM Modes}

The tilt functions of the lowest-order
(highest-frequency) undriven HFGM modes have a
single very narrow peak near the inner edge of
the disk. These modes are very localized spiral
corrugations of the inner disk. The lowest-order
undriven HFGM modes precess in the prograde
direction with precession frequencies comparable
to the gravitomagnetic precession frequency
$\omega_{\rm gm,i} = B/x_{i}^6$ at the inner
edge of the disk.

The damping rate of a mode is determined by the
radial spacing $\delta$ between the disk
corrugations. The damping rates $-\sigma$ of the
highest-frequency undriven HFGM modes are given
fairly accurately by $-\sigma \approx
 \pi^2 x_{\rm max}/\delta^2$, where $x_{\rm
max}$ is the position of the maximum of the tilt
function for the mode. Our numerical solutions
show that for the dozen or so highest-frequency
HFGM modes, $\delta$ decreases with increasing
mode number, but not smoothly, whereas the
radial width $\Delta$ of the tilt function
increases approximately linearly with increasing
mode number for $m>2$ up to at least $m \sim 14$.

With increasing mode number, the frequencies of
the HFGM modes at first decrease steadily. The
dozen or so lowest-order HFGM modes have
precession frequencies that range from $\approx
\omega_{\rm gm,i}$ for the $m=1$ mode to
$\sim0.5\,\omega_{\rm gm,i}$ for the $m=14$ mode.
These modes are underdamped, with $Q$ values that
range from 49 for the $m=1$ mode to 2.3 for the
$m=14$ mode. The damping rates of the HFGM modes
are sensitive to the viscosity in the disk. If
the viscosity parameter $\alpha$ (see \S~2) is
0.5 rather than 0.05, for example, the $Q$ of
the fundamental HFGM mode becomes 22 rather than
49.

With further increases in the mode number, the
HFGM modes become overdamped and their precession
frequencies eventually approach $\omega_{\rm
crit}$ asymptotically from above. The
highest-order HFGM modes are very
short-wavelength spiral corrugations that extend
from the radius $r_{\rm i}$ of the inner edge of
the disk only to $\sim10\,r_{\rm i}$. With
increasing mode number, the shapes of the HFGM
modes gradually approach the asymptotic shape of
the highest-order LFGMR modes. Like the
highest-order (highest-frequency) LFGMR modes,
the highest-order (lowest-frequency) HFGM modes
have precession frequencies $\sim \omega_{\rm
crit}$, which is $\sim\,1$~Hz for a compact
object of solar mass, and are very strongly
damped.

The warp functions, precession frequencies, and
damping rates of the modes in the HFGM family
depend on the boundary condition at the inner
edge of the disk. They are, for example,
affected by whether there is a driving torque
acting on the inner edge. Such a torque can
excite a variety of HFGM modes, in addition to
the free HFGM modes, depending on its time
dependence and its components parallel and
perpendicular to the angular momentum of the gas
in the disk. A torque that oscillates with a
given frequency and acts on the inner edge of
the disk can always resonantly excite the
appropriate driven HFGM mode that has the same
frequency, if the torque is strong enough.

The properties of the HFGM modes are only
slightly affected if the Newtonian gravitational
potential is replaced by a pseudopotential that
mimics the steeper effective potential at small
radii in general relativity. The HFGM modes are
also almost independent of the outer boundary
condition and the radius of the outer boundary.

The high-frequency HFGM modes may be involved in
some of the rapid X-ray variability and
high-frequency QPOs observed in some accreting
neutron stars and black hole candidates. The
highest possible frequency of the fundamental
(lowest-order, highest-frequency) HFGM mode is
the gravitomagnetic precession frequency of gas
at the radius $R_{\rm isco}$ of the innermost
stable circular orbit around the compact object.
For a compact object with mass $M$ and
dimensionless angular momentum $j \equiv cJ/GM^2
\ll 1$, $R_{\rm isco} \approx 6GM/c^2$ and hence
the Lense-Thirring precession frequency of the
innermost stable circular orbit is
 \begin{equation}
 \nu_{\rm gm,isco} \approx
   30\,\left(\frac{j}{0.2}\right)
   \left(\frac{2\msun}{M}\right)\;{\rm Hz}\;.
 \end{equation}
 The largest value of $j$ expected for neutron
stars with spin frequencies $\sim\,$300~Hz is
$\sim\,$0.2. Hence, for a $2\msun$ neutron star
with a radius smaller than that of the innermost
stable circular orbit and a spin frequency
$\sim\,$300~Hz, the {\em highest possible\/}
frequency of the fundamental HFGM mode would be
$\sim\,$30~Hz. The dozen lowest-order HFGM
modes, which are weakly damped, would then have
frequencies ranging from $\sim\,$30~Hz down to
$\sim\,$10~Hz. For a $6\msun$ black hole with $j
\sim 0.2$, the {\em highest possible\/}
frequency of the fundamental HFGM mode would be
$\sim\,$10~Hz and the dozen lowest-order HFGM
modes would have frequencies ranging from
$\sim\,$10~Hz down to $\sim\,$5~Hz.

As mentioned in \S~1, Stella \& Vietri (1998)
have recently suggested that gravitomagnetic
precession of the inner disk may be responsible
both for the broad bumps that are observed in
the power spectra of some atoll sources between
20 and 40~Hz, and for the horizontal branch QPO,
which is observed in the Z sources and appears
as a strong, narrow peak that varies in frequency
from 15 to 65~Hz. They proposed that these power
spectral features are caused by gravitomagnetic
precession of gas at the same radius as the gas
in circular Keplerian orbit that is thought to
generate the kilohertz QPOs (see Strohmayer et
al.\ 1996; Miller et al.\ 1998), which
have frequencies ranging from 900 to 1200~Hz in
the sources they considered.

In the picture proposed by Stella \& Vietri
(1998), the gravitomagnetic precession frequency
$\nu_{\rm gm}$, the frequency $\nu_{\rm K}$
of the Keplerian-frequency kilohertz QPO, and the
dimensionless angular momentum $j$ of the star
are related by
 \begin{equation}
 \nu_{\rm gm} =
      ({GM4\pi}/{c^3})\,j\nu_{\rm K}^2\;.
 \end{equation}
 Solving this relation for $I_{45}$, the moment
of inertia of the star in units of
$10^{45}$~\gcmsq, yields
 \begin{equation}
 I_{45}\left(\frac{\msun}{M}\right) =
 2\left(\frac{\nu_{\rm gm}^{}}
             {40_{}~\Hz}\right)
  \left(\frac{300~\Hz}{\nu_{\rm K}}\right)
  \left(\frac{1.2~\kHz}{\nu_{\rm K}}\right)^2\;.
 \end{equation}
 Computations of the {\em maximum\/} values of
$I_{45}\msun/M$ for modern neutron star equations
of state are very close to unity. For example,
for the FPS equation of state the maximum value
of $I_{45}\msun/M$ is 0.87 and occurs for a
1.55\msun\ star; for the UU equation of state the
maximum is 1.05 and occurs for a 2.1\msun\ star
(M.\,C. Miller, personal communication). Even for
the obsolete and unrealistically stiff equations
of state L and M, the maximum values of
$I_{45}\msun/M$ are $\lta\,$2.

These results show that the frequencies of the
bumps seen in the power spectra of the atoll
sources are about twice as high as the highest
precession frequencies predicted by modern
neutron star models, as Stella \& Vietri
themselves noted. Hence, in order to explain the
bumps in the atoll source power spectra by
gravitomagnetic precession of the inner disk, the
precessing HFGM corrugation would have to
generate X-ray brightness oscillations with a
frequency equal to twice the precession frequency
(i.e., at the second harmonic) or $I_{45}/M$ for
these neutron stars would have to be twice as
large as the largest values given by modern
neutron star equations of state.

The maximum frequencies of the horizontal branch
QPOs are $\sim\,$50\% higher than the maximum
frequencies of the bumps in the atoll source
power spectra and are therefore a factor
$\sim\,$3 higher than the highest precession
frequencies predicted by modern neutron star
equations of state, as Stella \& Vietri also
noted. For a detailed comparison of all the
available data on the atoll sources with the
predictions of the gravitomagnetic precession
model, see Psaltis et al.\ (1998).

In addition to resolving whether the frequencies
predicted by the gravitomagnetic precession
model can be reconciled with the frequencies of
the features observed in the power spectra of
neutron stars and black hole candidates, the
excitation, damping, and visibility in X-rays of
the low-order HFGM modes will have to be explored
further before they can be regarded as a promising
explanation for these features. Our results show
that these modes are not excited by the
radiation-warping torque. The two lowest-order
HFGM precession modes involve vertical motions
of the inner edge of the disk. This offers a
possible way of exciting them. In order to
excite steadily either of these two modes, a
torque acting on the inner edge of the disk
would have to match the variation of their
phases with azimuth as well as their
frequencies. On the other hand, coupling of the
vertical motion to the radiation field, magnetic
field, or surface of a neutron star may instead
damp these modes.

The two lowest HFGM modes might also be excited
by repetitive impulsive disturbance of the inner
disk. If, however, such disturbances affect the
inner portion of the disk as well as its
innermost edge, they may excite the broad
spectrum of underdamped HFGM modes, rather than
one or two such modes, producing a broad spectrum
rather than a QPO. Resonant excitation of the
$m>2$ HFGM modes appears difficult, because they
do not disturb the inner edge of the disk. In
order to excite these modes steadily, the driving
force would have to match closely the extremely
rapid variation of their phases with radius and
azimuth, as well as their frequencies.

The highest-order HFGM modes do not appear
promising as a mechanism for producing
quasi-periodic oscillations in X-ray brightness,
because they are very tightly wound spiral
corrugations within a narrow annulus centered at
a radius $\sim\,$5 times larger than the radius
of the inner edge of the accretion disk and do
not affect the accretion disk at its inner edge.
On their face, the lowest-order HFGM modes
appear more capable of causing oscillations in
the X-ray emission, for example by modulating
the accretion onto the compact object.

A more complete exploration of the detailed
properties of the high-order radiation-warping
modes,\break LFGM modes, LFGMR modes, and HFGM
modes will be presented elsewhere.

\acknowledgements

We are grateful to Philip Maloney, Cole Miller,
and Dimitrios Psaltis for many helpful comments,
and to Cole Miller for computing the moments of
inertia used in \S~4. This work was supported in
part by NSF grants AST~93-15133 and AST~96-18524
and by NASA grant NAG~5-2925.

\end{document}